\documentclass[12pt,twoside]{article}
\usepackage{fleqn,espcrc1}
\usepackage{amssymb}
\usepackage{graphics,graphpap}
\usepackage{graphicx}
\usepackage[figuresright]{rotating}

\newcommand{\AmS}{{\protect\the\textfont2
  A\kern-.1667em\lower.5ex\hbox{M}\kern-.125emS}}

\title{A Non-Perturbative Chiral Approach for Meson-Meson Interactions 
\thanks{Talk given at PANIC99, Uppsala (Sweden), June 10-16, 1999.}}
\author{J.A. Oller, E. Oset
\address{Departamento de F\'{\i}sica Te\'orica \\
Universidad de Valencia, 46100 Burjassot (Valencia), Spain} 
\thanks{Work partially supported by DGICYT under contract PB96-0753 and by the
EEC-TMR Program-Contrac No. ERBFMRX-CT98-0169. J.A.O. acknowledges financial
support from the Generalitat Valenciana.}
F. Guerrero
\address{Centro Universitario Estema\\ Parque Tecnol\'ogico, 46980 Valencia,
Spain}
J.R. Pel\'aez
\address{Departamento de F\'{\i}sica Te\'orica\\
Universidad Complutense, 28040 Madrid, Spain}}                    

\begin{document}

% typeset front matter
\maketitle

\begin{abstract}
A non-perturbative method \cite{prl} which combines constraints from chiral
symmetry breaking and coupled channel unitarity is used to describe 
meson-meson interactions up to $\sqrt{s}\lesssim 1.2$ GeV, extending in this way
the range of applicability of the information contained in Chiral Perturbation
Theory ($\chi PT$) \cite{xpt}, since this perturbative series is typically 
restricted to $\sqrt{s}\lesssim 500$ MeV. The approach uses the 
$\mathcal{O}(p^2)$ and $\mathcal{O}(p^4)$ $\chi PT$ Lagrangians. The seven 
free parameters resulting from the $\mathcal{O}(p^4)$
Lagrangian are fitted to the experimental data. The approach makes use of the
expansion of $T^{-1}$ instead of the amplitude itself as done in $\chi PT$.
The former expansion is suggested by analogy with the effective range
approximation in Quantum Mechanics and it appears to be very useful. The
results, in fact, are in good agreement with a vast amount of experimental
analyses \cite{guerrero,prd}.

The amplitudes develop poles corresponding to the
$f_0(980)$, $a_0(980)$, $\rho(770)$, $K^*(890)$, the octet contribution to the
$\phi$, $f_0(400-1200)\equiv \sigma$ and $\kappa$ \cite{prd}. The total and 
partial decay widths of the resonances are also well reproduced.

\end{abstract}

\section{Introduction} \label{sec:intro}
$\chi PT$ is the low energy effective theory of the strong interactions. It is
given as a power expansion of the external four-momenta of the
pseudo-Goldstone bosons $\pi$, $K$ and $\eta$ on the scale $\Lambda_{\chi
PT}\approx$1 GeV. As a result, the expansion is typically valid up to
$\sqrt{s}\lesssim$500 MeV. However, the constraints coming from the
spontaneous/explicit chiral symmetry are not restricted to the low energy
region \cite{zahed}. In this work, we present a way of resummation of
the $\chi PT$ series that in fact can be applied to any other system whose
dynamics can be described by low energy chiral Lagrangians. We describe the
successfull application of such approach to meson-meson interactions
which are well reproduced up to $\sqrt{s}\lesssim$1.2 GeV.

\section{Formalism}

Let us a consider a partial wave amplitude $T$ with definite isospin ($I$).
We use a matrix formalism in order to deal with coupled channels. In this way
$T$ will be a matrix whose element $ij$ represents the scattering of
$i\rightarrow j$ with angular momentum $L$ and isospin $I$. If we consider
only two body intermediate states unitarity with
coupled channels reads in our normalization:

\begin{equation}
\label{uni}
\hbox{Im}T^{-1}=\rho
\end{equation}
where $\rho$ is a diagonal matrix with elements $\displaystyle{\rho_i=
\frac{p_i}{8\pi\sqrt{s}}} \theta(s-(m_{1i}+m_{2i})^2$ with $p_i$ the center mass
three-momentum, $m_{1i}$ and $m_{2i}$ are the masses of the particles in the
state $i$ and $\theta(x)$ is the usual Heaviside function. Eq. (\ref{uni}) is a 
well known result and is the basis
of the $K$ matrix formalism since all the dynamics is embodied in Re$T^{-1}$
which is $K^{-1}$. The former equation shows clearly that, when considering 
$T^{-1}$, unitarity is exactly satisfied with two body intermediate
states.

From the $\chi PT$ expansion of $T=T_2+T_4+\mathcal{O}(p^6)$, where $T_2$ and
$T_4$ are the $\mathcal{O}(p^2)$ and $\mathcal{O}(p^4)$ contributions
respectively, we
work out the expansion of $T^{-1}$. In this way we will obtain our approach for
the $K$ matrix (or Re$T^{-1}$).

\begin{eqnarray}
\label{t-1}
T^{- 1} &=& \left[T_2+T_4+...\right]^{-1}=
T_2^{- 1}\cdot [1 + T_4 \cdot T_2^{- 1}+...]^{- 1}\nonumber \\ 
&=& T_2^{- 1}\cdot [1 - T_4 \cdot T_2^{- 1}+...]=T_2^{-1}\cdot [T_2-T_4]\cdot
T_2^{-1}
\end{eqnarray}
Inverting the former result, one obtains:

\begin{eqnarray}
\label{t}
T&=&T_2\cdot \left[T_2-T_4 \right]^{-1}\cdot T_2 \nonumber \\
K&=&T_2\cdot \left[T_2-\hbox{Re}T_4 \right]^{-1}\cdot T_2
\end{eqnarray} 

\section{$\pi\pi$ and $K\bar{K}$ coupled amplitudes}

In \cite{guerrero} we study the $(I,L)=(0,0),(1,1)$ and $(2,0)$ partial waves.
To make use of eq. (\ref{t}) one needs the lowest and next to leading order $\chi
PT$ amplitudes. In our case the $\pi\pi\rightarrow \pi\pi$ and
$\pi\pi\rightarrow K\bar{K}$ are taken from \cite{UM} and the
$K\bar{K}\rightarrow K\bar{K}$ is also given in \cite{guerrero}. Our amplitudes
depend on six parameters $L_1$, $L_2$, $L_3$, $L_4$, $L_5$ and $2 L_6+L_8$ which
are fitted to the elastic $\pi\pi$ $(I,L)=(0,0)$ and $(1,1)$ phase shifts.

\begin{figure}
\begin{minipage}[t]{0.4\textwidth}
\includegraphics[width=\textwidth,angle=-90]{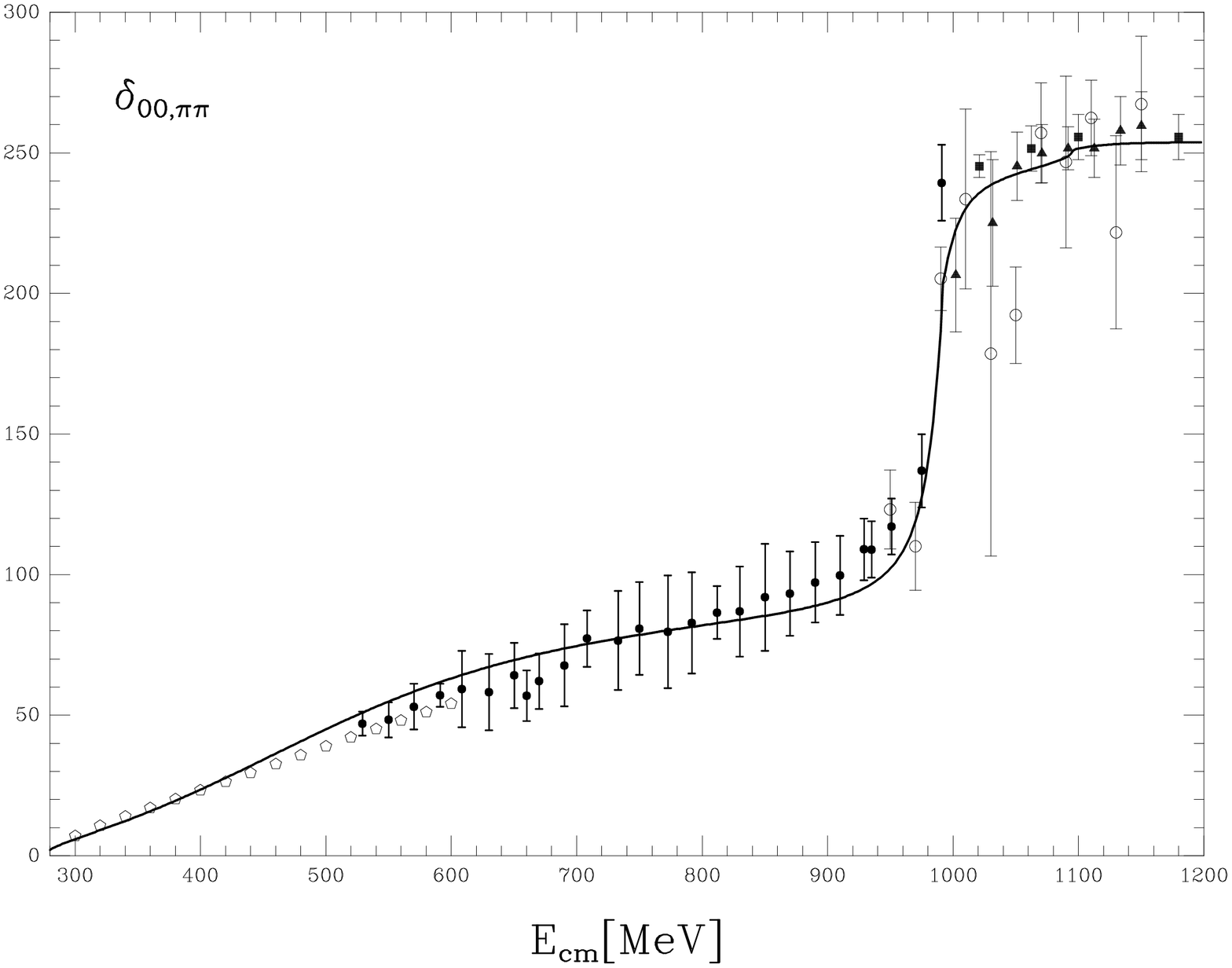}
\caption{\label{fig1} Elastic P-wave $\pi\pi$ phase shifts. References in 
\cite{guerrero}.}
\end{minipage}
\hfill
\begin{minipage}[t]{0.4\textwidth}
\includegraphics[width=\textwidth,angle=-90]{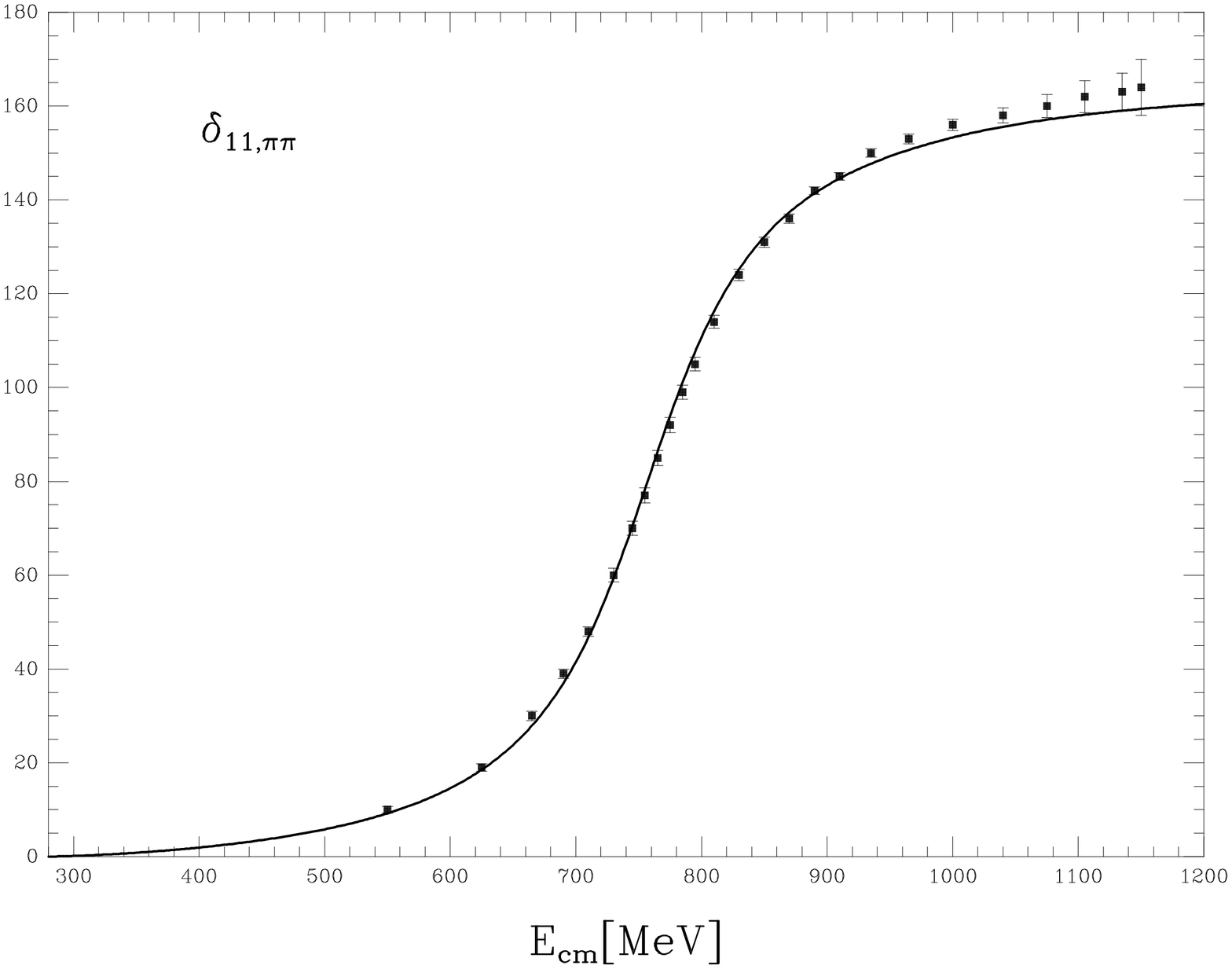}
\caption{\label{fig2} Elastic S-wave $I=0$ $\pi\pi$ phase shifts. References 
in \cite{guerrero}.}
\end{minipage}
\end{figure}

In the following table we show the resulting values for the $L_i$ coefficients
comparing them with the $\chi PT$ values.

\begin{table}[ht]
\caption{$L_i$ coefficients.}
\begin{tabular}{ccc}
\hline
& Fit&$\chi PT$ \\
\hline
$L_1\,10^{3}$& $0.72^{+0.03}_{-0.02}$ & $0.4\pm 0.3$ \\
$L_2\,10^{3}$& $1.36^{+0.02}_{-0.05}$ & $1.4\pm 0.3$ \\
$L_3\,10^{3}$& $-3.24\pm 0.04$ & $3.5 \pm 1.1$ \\
$L_4\,10^{3}$&$0.20\pm 0.10$ & $-0.3\pm 0.5$\\
$L_5\,10^3$&$0.0^{+0.8}_{-0.4}$&$1.4\pm 0.5$\\
$(2 L_6+L_8)\,10^3$&$0.00 ^{+0.26}_{-0.20}$&$0.5\pm 0.7$\\
\hline
\end{tabular}
\end{table}

With the former values for the $L_i$ couplings we also calculate other
scattering parameters in good agreement with experiment. 

It is worth to indicating that from eq. (\ref{t}) the $\chi PT$ expansion 
is recovered for low energies up to $\mathcal{O}(p^4)$. In this way, we also 
calculate in \cite{guerrero} the scattering lengths with values in 
agreement with $\chi PT$ and experiment.

\section{S and P-wave meson-meson scattering amplitudes}

In \cite{prd} we thoroughly study the meson-meson interactions for $L=0$ and 1
making use of eq. (\ref{t}). However, in this case there are a lot of
channels whose $\chi PT$ $T_4$ amplitudes have not been calculated yet
\footnote{Even more, when 
this work was done the $K\bar{K}\rightarrow K\bar{K}$ amplitudes were 
not calculated.}. The calculation of the $T_4$ although straightforward is
cumbersome. As a result, we approximate the $T_4$ amplitude as in \cite{prd}

\begin{equation}
\label{t4}
T_4\approx T_4^P+T_2\cdot g(s) \cdot T_2
\end{equation}
where $T_4^P$ is the polynomial part of the amplitude which is essential for the
vector channels (this is another way to see Vector Meson Dominance \cite{grupo})
and $T_2\cdot g(s) \cdot T_2$ takes into account unitarity in coupled channels,
mostly important for the scalar channels. The $g(s)$ function is a diagonal
matrix whose elements are the loop integral with two meson propagators:

\begin{equation}
g_{i}(s) = i \int \frac{d^4 q}{(2 \pi)^4} \;
\frac{1}{q^2 - m^2_{1 n} + i \epsilon} \;
\frac{1}{ (P - q)^2 - m^2_{2 n} + i \epsilon}
\label{g}
\end{equation} 

\vspace{-0.5cm}

We regularize it making use of a cut-off $q_{max}$.

With respect to a full $\chi PT$ calculation we are neglecting in eq. 
(\ref{t4}) the 
tadpole contribution (which numerically
usually results to be small \cite{UM}) and the unphysical cuts (which
correspond to singularities away from the physical region and hence they give 
rise to soft contributions which we reabsorb in the $L_i$ couplings).

After inserting eq. (\ref{t4}) with eq. (\ref{g}) in eq. (\ref{t}) we obtain the
final expression for the $T$ matrix as in \cite{prd}. We reproduce in that work
a vast amount of experimental data (phase shifts and inelasticites) for the S
and P-wave meson-meson scattering. We also study the mass, widths and partial 
decay widths of the resonances (poles) present in our amplitudes:

\begin{table}[ht]
\caption{Masses and partial widths in MeV.}
\begin{tabular}{|c|c|c|c|c|c|c|}
\hline
  $ \begin{array}{c} {\rm Channel}  \\ (I,J)       \\ \end{array} $ 
&  Resonance 
& $ \begin{array}{c}  {\rm Mass }   \\ {\rm from \, pole}  \\ \end{array} $
& $ \begin{array}{c}  {\rm Width}   \\ {\rm from \, pole}  \\ \end{array} $
& $ \begin{array}{c}  {\rm Mass}   \\ {\rm effective} \\ \end{array} $
& $ \begin{array}{c}  {\rm Width}  \\ {\rm effective} \\ \end{array} $ 
& $ \begin{array}{c}  {\rm Partial}\\ {\rm Widths}\\ \end{array} $ \\ 
\hline  
$(0,0)$
& $\sigma $
& $442$
& $454$
& $\approx$ $600$ 
& $very$ $large$
&$\pi\pi - 100\%$\\
\hline
$(0,0)$
& $f_0(980)$
& $994$
& $28$
& $\approx$ $980$ 
& $\approx$ $30$
& $\begin{array}{c} \pi\pi -65\%\\ K \bar{K} - 35\% \\ \end{array}$ 
\\ 
\hline
$(0,1)$
& $\phi(1020)$
& $980$
& $0$
& $980$
& $0$
& 
\\
\hline
$(1/2,0)$
& $\kappa$
& $770$
& $500$
& $\approx$ $850$
& $very$ $large$
& $K \pi - 100\%$
\\
\hline
$(1/2,1)$
& $K^*(890)$
& $892$
& $42$
& $895$
& $42$
& $K \pi -  100\% $
\\
\hline
$(1,0)$
& $a_0(980)$
& $1055$
& $42$
& $980$
& $40$
& $\begin{array}{c}\pi\eta - 50 \% \\ K \bar{K} -  50\% \\ \end{array}$
\\ 
\hline
$(1,1)$
& $\rho(770)$
& $759$
& $141$
& $771$
& $147$
& $\pi\pi - 100\%$
\\
\hline
\end{tabular}
\end{table}

\section{Conclusion}

We have presented a method of resummation of the $\chi PT$ series based in the
expansion of $T^{-1}$. In this way unitarity is fulfilled to all orders and
resonances are well reproduced. The method is rather general and could be
applied to any system whose dynamics is described by chiral Lagrangians. We have
applied it successfully to describe the S and P-wave meson-meson amplitudes
giving rise to the resonances: $f_0(980)$, $a_0(980)$, $\rho(770)$, $K^*(890)$, the octet contribution to the
$\phi$, $f_0(400-1200)\equiv \sigma$ and $\kappa$.

\end{document}